\documentclass[aps,prd,twocolumn,superscriptaddress]{revtex4}
\usepackage{graphicx}
\usepackage{xcolor}
\usepackage{amsmath,amssymb,latexsym}
\usepackage{hyperref}
\usepackage{txfonts}
\usepackage{ulem}

\begin{document}

\title{Stochastic gravitational-wave background from spin loss of black holes}
\author{Xi-Long Fan}
\email{fanxilong@outlook.com}
\affiliation{Department of Physics and Mechanical and Electrical Engineering, Hubei University of Education, Wuhan 430205, China}
\author{Yanbei Chen}
\email{yanbei@caltech.edu}
\affiliation{Theoretical Astrophysics and the Burke Institute for Theoretical Physics 350-17, California Institute of Technology, Pasadena, CA 91125, USA}

\begin{abstract}
Although spinning black holes are shown to be stable in vacuum in general relativity, exotic mechanisms 
have been speculated to convert the spin energy of black holes into gravitational waves.  Such waves may be very weak in amplitude, since the spin-down could take a long time, therefore a direct search may not be feasible.  
We propose to search for the stochastic red gravitational-wave background associated with the spin-down, and we relate the level of this background to the formation rate of spinning black holes from the merger of binary black holes, as well as the energy spectrum of waves emitted by the spin-down process.    We argue that current LIGO-Virgo observations are not inconsistent with the existence of a spin-down process, as long as it is slow enough. On the other hand, the background may still detectable as long as a moderate fraction of spin energy is emitted within Hubble time.    This stochastic background could be one interesting target of  next generation GW detector network, such as  LIGO Voyager, and could  be extracted  from total stochastic background. 
\end{abstract}
\maketitle

\section{Introduction: Motivation and Key Assumptions}
\label{sec:intro}
  Spinning black holes are known to contain energy that can be extracted --- even with classical physical processes (e.g. Penrose process \cite{1971NPhS..229..177P} and Blandford-Znajek process \cite{1977MNRAS.179..433B}).  The area theorem dictates a limit of extraction energy $\Delta E \le M-M_{\rm irr}$, given by the difference between the mass $M$ of the black hole, and its irreducible mass $M_{\rm irr}$ defined by:
\begin{equation}
M_{\rm irr} =  \sqrt{\frac{1+\sqrt{1-(a/M)^2}}{2}} M ,
\end{equation} 
where $a$ is the spin of the  black hole~\cite{PhysRevLett.25.1596}.
This extraction is believed to be powering highly energetic astrophysical processes (e.g. \cite{1977MNRAS.179..433B}).  More mathematically speaking, near spinning black holes, perturbations which enter the horizon that are co-rotating with the black hole, with a slower angular velocity, carries negative energy down the black hole, thereby transferring positive energy toward infinity.  Such an effect is commonly referred to as {\it superradiance}~\cite{2015LNP...906.....B,press1972floating}.  

In this paper, we will explore the possible existence of a stochastic gravitational-wave background due to black-hole superradiance. 

\subsection{Possible Spin-Down Mechanisms}

Superradiance causes perturbations to be  unstable in some  cases: (i)  photons acquire mass due to dispersion when propagating through plasma~\cite{teukolsky1974perturbations,conlon2017radionovas}, (ii) for a massive scalar/vector field~\cite{detweiler1980klein}, such as the axion and possibly other bosons~\cite{2011PhRvD..83d4026A,2015PhRvD..91h4011A,east2017superradiant,2017PhRvL.119m1101B,PhysRevD.96.064050,2018arXiv180700043E} , and (iii) if Kerr black hole transitions into an ultracompact object, or a gravastar  ~\cite{chirenti2008ergoregion,cardoso2008ergoregion}.  If a spinning black hole/gravastar were to form anyway, then this linear instability should lead to a Spin-Down (SD).  In cases (ii) and (iii), this will lead to the conversion of spin energy into gravitational waves, through the re-radiation of gravitational waves by an axion or boson  cloud in (ii), and through direct emission of gravitational waves in (iii).   

More specifically, such emissions from mechanism (ii) mentioned above was proposed by Arvanitaki and Dubovsky~\cite{2011PhRvD..83d4026A} as a way to search for axions; the emission mechanism was later studied numerically by East and Pretorius~\cite{east2017superradiant}; more recently, it was proposed to search for this type of emission in gravitational-waves that follow  binary black-hole mergers~\cite{2015PhRvD..91h4011A,2017PhRvD..95d3001A,2017PhRvD..96c5019B}.  Furthermore, the stochastic gravitational-wave background  that arise from various axion spins and masses have been studied  extensively in by Brito et al.~\cite{2017PhRvL.119m1101B,PhysRevD.96.064050}.

The mechanism (iii) has been studied qualitatively by Chirenti and Rezzolla ~\cite{chirenti2008ergoregion} and Cardoso et al.~\cite{cardoso2008ergoregion}, as arguments that long-lived spinning gravastars should not exist.   

Inspired by these individual spin-down models,  we believe there is enough motivation to consider more generic parametrized models for the spin-down of Kerr black holes, and discuss its detectability by current and future gravitational-wave detectors.

\subsection{Observational Constraints}

The possibility of spin-down  does not necessarily mean that rapidly spinning Kerr black holes, or spinning gravastars, do not exist in nature, because (i) for non-isolated black holes, angular momentum carried away by the spin-down mechanism can be balanced by accretion, and (ii)  the spin-down rate can be low and the spin angular momentum can take a long time to radiate away.

Significant spins of  stellar-mass black holes in X-ray binaries and supermassive black holes  at the  center of galaxies have been estimated  by measuring  properties of the accretion disk  through continuum fitting method and  x-ray relativistic reflection method, respectively (see e.g., Ref.~\cite{2013arXiv1312.6698N} for a review on this subject).   Angular momentum carried by the accretion flow can presumably balance the spin-down mechanism that exist for such systems, and continue to spin up the black hole.  This will give rise to an additional gravitational-wave background, which we do not study in this paper.  However, such background from in the particular type of superradiance has been studied by Baryakhtar {\it et al.}~\cite{2017PhRvD..96c5019B}.

On the other hand, in the binary black-hole merger events detected by  Advanced LIGO and Advanced Virgo,  the individual merging black holes may all have low or zero spins ~\cite{2017arXiv170601812T,2017arXiv171105578T,2017PhRvL.119n1101A}.   This is at least consistent with a spin-down time scale that is  at least a sizable fraction of Hubble time, and may even be used as an evidence of spin-down from the original formations of the individual black holes. 

Finally, significantly spinning black holes do form due to binary black hole mergers, as so far have been detected by Advanced LIGO and Virgo~\cite{2016PhRvL.116f1102A,2016PhRvL.116x1103A,2017arXiv170601812T,2017arXiv171105578T,2017PhRvL.119n1101A}, which estimates a local merger rate of 12-213 \,${\rm Gpc}^{-3}{\rm yr}^{-1}$ \cite{2017arXiv170601812T}.     We will use this formation channel to provide the source for spin-down emission.

\subsection{BBH Mergers as source of spin energy}

For equal-mass binaries,  the final black hole has $a/M \approx 0.7$, with around 7\% of its rest mass stored in spin energy, which is larger than the gravitational-wave energy radiated during the Inspiral, Merger and Ringdown (IMR) processes combined, which is roughly 5\%~\cite{PhysRevLett.95.121101}. In this way, if a non-trivial fraction of the spin energies of these newly formed black holes can be radiated away in the form of gravitational waves during Hubble time, such radiation will form a non-trivial, or even stronger, gravitational-wave background.   
 Since the IMR background is already plausible for detection in second-generation detector networks \cite{2016PhRvL.116m1102A}, and Advanced LIGO will be  updated to Advanced LIGO + (AL+),  LIGO Voyager (Voyager) \footnote{LIGO Document T1500293-v8 and LIGO Document T1700231-v3},  this additional background is well worth studying.

\begin{figure}
\includegraphics[width=0.45\textwidth]{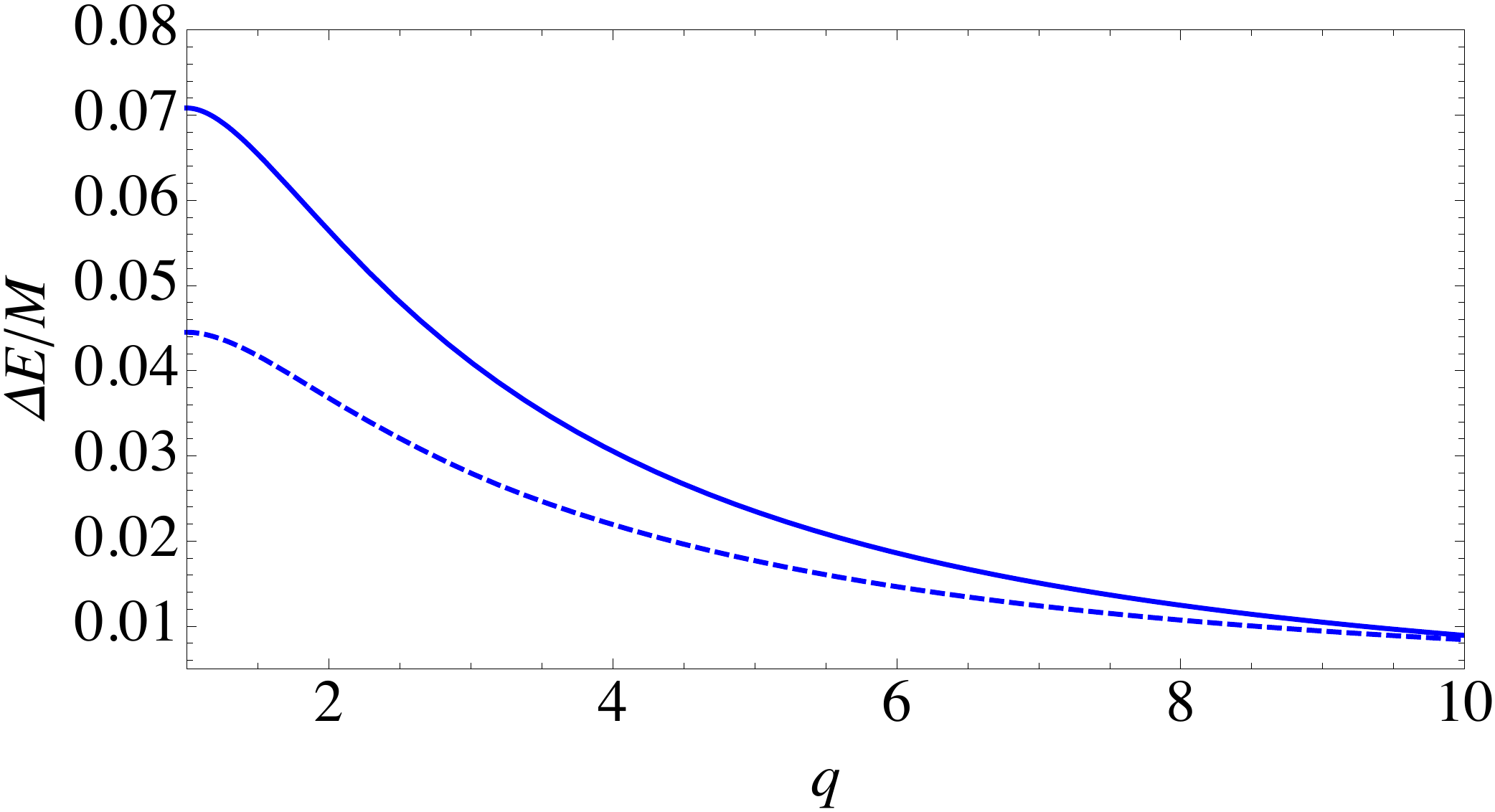}
\caption{Energy stored in the spin of the final merger product (solid line), in comparison with energy radiated during the entire coalescence (dashed line). \label{fig:energy}}
\end{figure}

\subsection{Summary of Key Assumptions}

Here we list our key assumptions --- based on which we shall  argue that a spin-down gravitational-wave background is observationally relevant: (i)  at least a moderate fraction of the spin energy of black holes are lost over time, (ii) at least a moderate fraction of the energy loss is via gravitational waves, (iii) the spin-down takes place within the Hubble time.  As we have argued earlier in this section, these assumptions, while speculative, are not only consistent with  the low  spins of individual merging black holes estimated with GW observations and the rapid spins of black holes in  X-Ray binary systems,  but has also been motivated by particular models considering axions around Kerr black holes~\cite{2011PhRvD..83d4026A,east2017superradiant,2017PhRvL.119m1101B,PhysRevD.96.064050} and rotating gravastars~\cite{chirenti2008ergoregion,cardoso2008ergoregion}.  

In addition to (i) --- (iii) above, we shall further assume: (iv) the spin-down gravitational-wave energy spectrum can be roughly captured by models  that will be proposed later in this paper. 


\subsection{Outline of the paper}

This paper is organized as follows. In Sec.~\ref{subsec:single}, we construct two models for the spin-down emission spectrum of a single binary black-hole merger remnant, a {\it Parametrized Gaussian Model} in which the central frequency and width are parametrized, and a {\it fiducial model}, in which we assume the emission at any given time is centered narrowly at the black hole's fundamental quasi-normal mode (QNM) frequency. In Sec.~\ref{subsec:stoch}, we assemble the single-remnant spectrum into a stochastic background.    In Sec.~\ref{subsec:snr}, we compute the signal-to-noise ratio of the spin-down stochastic background, when taking correlations between output from pairs of detectors; in Sec.~\ref{subsec:extract},
we formulate a Fisher-Matrix approach to estimate how well the SD background can be extracted, with results shown in Sec.~\ref{subsec:results}. Finally, we summarize our conclusions and discuss subtleties in our results in Sec.~\ref{sec:conclusion}.

\section{The Stochastic Background}
\label{sec:bkg}

Let us estimate the magnitude of the stochastic background due to spin down, by first estimate the emission of a single binary black-hole remnant in Sec.~\ref{subsec:single}, and then synthesize the stochastic background as the superposition of all the remnants, in Sec.~\ref{subsec:stoch}. 

\subsection{Emission from a single remnant} 
\label{subsec:single}

 For a binary of Schwarzschild black holes  with masses $M_{1,2}$ and mass ratio $q\equiv M_1/M_2$, the spin $a_0$ and final mass $M_0$ of the new-born  merged black hole has the following dependence on the symmetric mass ratio $\eta \equiv M_1 M_2/(M_1+M_2)^2$~\cite{rezzolla2008final}: 
\begin{equation}
 \frac{a_0}{M_0} = 2\sqrt{3} \eta -3.454\eta^2+2.353\eta^3.
\end{equation}
Assuming that all spin energy is radiated as gravitational waves,  we obtain the spin-down energy :
\begin{equation}
\Delta E^{\rm SD}_{\rm tot} = M -M_{\rm irr} = \left(1-\sqrt{\frac{1+\sqrt{1-\chi^2}}{2}}\right)M, 
\end{equation} 
where  $\chi \equiv a/M$ is the dimensionless spin. As indicated by  Fig.~\ref{fig:energy}, for comparable masses (with mass ratio  $q$ close to unity),  $\Delta E^{\rm SD}_{\rm tot}$  is always significantly larger than $\Delta E^{\rm IMR}_{\rm tot}$~\cite{PhysRevD.90.104004}.  

In the following, we shall make two different models for the frequency spectrum  $d E^{\rm SD}/df$.  The first assumes that the spectrum is a Gaussian,
\begin{equation}
\label{eq:gaussian}
\left(\frac{dE^{\rm SD}}{df}\right)_{\rm Gauss} = \frac{ \Delta E_{\rm tot}^{\rm SD}}{\sqrt{2\pi} f_c/q}\exp\left[-\frac{(f-f_c)^2}{2(f_c/Q)^2}\right]\,,
\end{equation}
where $f_c$ is the central frequency, which we will prescribe to be $  f_c = \beta/M$, with $\beta$ a (mass- and spin-independent) constant, $Q$ is a constant quality factor. We shall refer to this as the {\it Parametrized Gaussian Model}; note that $\beta$ and $Q$ are left as tunable parameters.  Such Gaussian approximation approach  is also adopted to investigate the gravitational wave background  from
core collapse supernovae, which spectrum is not very clear yet \cite{2010MNRAS.409L.132Z}.

\begin{figure}
\includegraphics[width=0.475\textwidth]{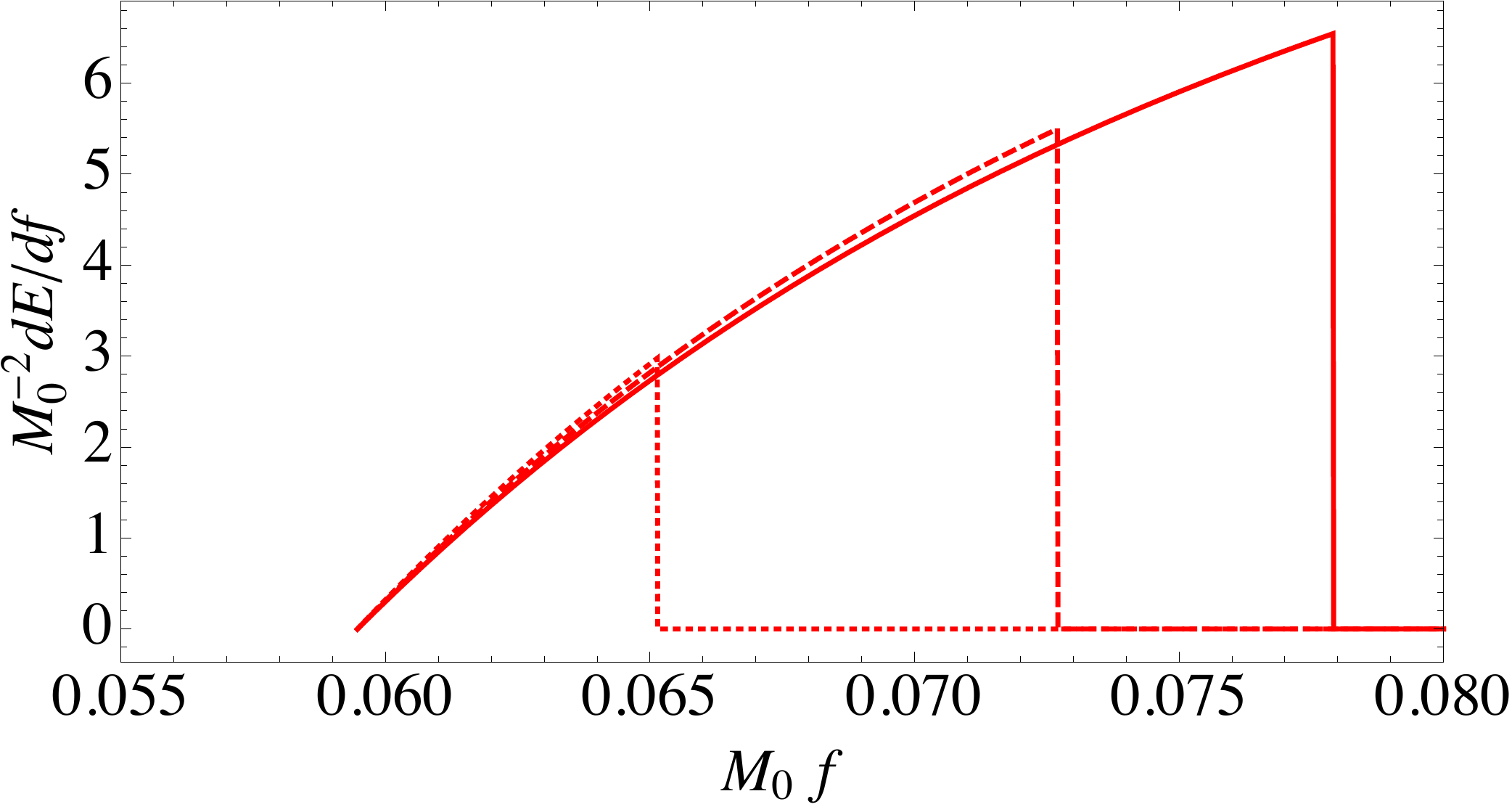}
\caption{Radiation spectrum during the spin decay process in the fiducial model, assuming that radiation is predominantly at the first QNM of the merger product, for mass ratio  $q=1$ (solid), 3 (dashed) and 10 (dotted).  For the alternative, Parametrized Gaussian Model, see Eq.~\eqref{eq:gaussian}.  \noindent \label{fig:spectrum}}
\end{figure}

As another model, let us assume that at any given moment, the emission is only at the lowest $l=m=2$ quasi-normal mode (QNM) frequency of the Kerr black hole,
\begin{equation}
f_{\rm QNM}(M,a) = M^{-1} F(a/M) \,,
\end{equation}
where $F$ is given by, e.g., Eq.~(4.4) of Ref.~\cite{1989PhRvD..40.3194E}. We shall refer to this as the QNM model, and it will be part of our {\it fiducial model}. The choice here is rather speculative; the bases for doing so is by noting that: (i) the (2,2) mode is usually the most radiated gravitational-wave mode, and (ii) the QNM frequency is a characteristic of the potential barrier for wave propagation outside of black holes, and has been seen as the peak of the frequency spectra of {\it echoes} if reflecting surfaces exist right outside the horizon.

Assuming that $M_{\rm irr}$ remains the same throughout the spin-down process, we obtain
\begin{equation}
\left(\frac{dE^{\rm SD}}{df}\right)_{\rm QNM} =\left[\frac{\partial M}{\partial \chi}\right]_{M_{\rm irr}} \Bigg/ \left[\frac{\partial f_{\rm QNM}}{\partial \chi}\right]_{M_{\rm irr}},
\end{equation}
where both $M$ and $f_{\rm QNM}$ are written  in terms of $M_{\rm irr}$ and $\alpha$ :
\begin{equation}
M( M_{\rm irr},\alpha) = M_{\rm irr} /\sqrt{1+\sqrt{1-\chi^2}}\,,
\end{equation}
and 
\begin{equation}
f_{\rm QNM}(M_{\rm irr},\alpha) =  \sqrt{1+\sqrt{1-\chi^2}}M_{\rm irr}^{-1} F(\chi).
\end{equation} 
For a new-born merged black hole with mass $M_0$ and dimensionless spin $\chi_0$, we first compute $M_{\rm irr}$, which remains fixed during the spin-down, then obtain both $f$ and $dE/df$ as functions of $\chi$, with $\chi$ decreasing from $\chi_0$ to 0.
In Fig.~\ref{fig:spectrum}, we plot $M_0^{-2}(dE^{\rm SD}/df)_{\rm QNM}$ as functions of $M_0 f$, for Kerr black holes that form from binaries with $q=$1, 3 and 10, with $\chi =$0.69, 0.54 and 0.26, respectively.  
 
\subsection{ Stochastic Background}  
\label{subsec:stoch}
By using knowledge about cosmology and binary black-hole merger rate throughout ages of the universe, the energy  spectrum of  the spin-down of the final black hole produced by a single binary merger  can be converted into the energy density spectrum of the stochastic background, which we express in terms of the energy density per logarithmic frequency band, normalized by  the closing energy density of the universe \cite{1999PhRvD..59j2001A,2016PhRvL.116m1102A}:
\begin{eqnarray}
&&\Omega_{\rm GW} (f_{\rm obs}) \equiv  {\rho_c}^{-1}  \left[{d\rho(f)}/{ d \log  f}\right]_{f=f_{\rm obs}} \nonumber\\
=&&\int d\theta \int_{0}^{z_{\rm max}} dz \frac{  f_{\rm obs} R_m(z,\theta) \displaystyle
\left[\frac{dE(f,\theta)}{df}\right]_{f=(1+z)f_{\rm obs}}}{(1+z)\rho_c H_0 E(\Omega_M,\Omega_\Lambda,z)}.\;
\label{Omega}
\end{eqnarray}
Here  we assume a family of sources parametrized by $\theta$ (e.g., masses and spins), with  $R_m(z,\theta)$  the event rate density  per $\theta$ volume, per co-moving volume at redshift $z$,  and 
\begin{equation} 
E(\Omega_M ,\Omega_\Lambda,z) =\sqrt{\Omega_M(1+z)^3+\Omega_\Lambda},
\end{equation}
 \cite{1999astro.ph..5116H}.  We use $H_0=70$ km s$^{-1}$ Mpc$^{-1}$, $\Omega_M=1-\Omega_\Lambda=0.28$ in this paper \cite{2013ApJS..208...20B}.  For each $z$, one can define
\begin{equation}
\mathcal{R}_m (z) = \int R_m (z,\theta)\,d\theta \,,\quad 
p(z,\theta) = R(z,\theta)/\mathcal{R}_{m} (z)
\end{equation}
with $\mathcal{R}_m(z)$ the total rate per unit co-moving volume at redshift $z$, and $p(z,\theta)$  the distribution density of source parameter $\theta$ at redshift $z$.

\section{Detectability} 

Let us now explore the detectability of the SD background, by first computing the signal-to-noise ratio including and excluding this background in Sec.~\ref{subsec:snr}, then proposing a Fisher-Matrix formlaism to estimate how well the SD background can be extracted in Sec.~\ref{subsec:extract}, and finally showing results in Sec.~\ref{subsec:results}.

\subsection{Signal-to-noise ratio}
\label{subsec:snr}

The optimal  signal-to-noise ratio (SNR) for the  the total   background energy density spectrum is given by
\begin{equation}
  {\rm SNR} =\frac{3 H_0^2}{10 \pi^2} \sqrt{2T} \left[
\int_0^\infty df\>
\sum_{i> j}
\frac{\gamma_{ij}^2(f)\Omega_{\rm GW}^2(f)}{f^6 S_h^i(f)S_h^j(f)} \right]^{1/2}\,,
\label{eq:snrCC}
\end{equation}
for a network of detectors $i,j=1,2,\cdots, n$, where $S_h^i(f)$ is the  one-sided strain noise
spectral density of  detector $i$; $\gamma_{ij}(f)$ is the {\it normalized
isotropic overlap reduction function} between the $i$ and $j$ detectors, and $T$ is the accumulated coincident observation time of detectors.  To  detect a stochastic background with $   90\%$ and $99.7\%$  confidence, the SNR should be larger than 1.65 and 3, respectively  \cite{1999PhRvD..59j2001A}.  Note that this SNR is only achievable when our template for the shape of $\Omega_{\rm GW}(f)$ is optimal.  

As we see from Eq.~\eqref{Omega},  the total energy  density spectrum $\Omega_{\rm GW}$  mainly depends  on  the merger rate   of one class of source $\mathcal{R}_m$,  source population properties (such as mass distribution)  $p(z,\theta)$  and the spectral  energy density of a single  source $dE/df$.  The detail effects of merger rate and source  mass distribution  are discussed in  \cite{2016PhRvL.116m1102A,2017PhRvL.118l1101A}.  These two ingredients  have weak effects on the background spectrum shape, especially  in the Advanced LIGO- Advanced Virgo network band 10-50 Hz, where  the spectrum is well approximated by a power law  $\Omega_{\rm GW} \sim f^{2/3}$  (see detail discussion and references in \cite{2013MNRAS.431..882Z,2016PhRvL.116m1102A}).   Note that,  the spectral  energy density $(dE/df)_{\rm IMR}$  of single  source adopted in most literature is  only the leading harmonic of the GW signal (e.g.\cite{2016PhRvL.116m1102A,2011PhRvD..84h4037A}),  which is reasonable for current ground detectors, since the overlap reduction function  modified  the most sensitive band to 10-50 Hz.  Our {\it fiducial} QNM model is constructed as follows:
(i) we assume $\mathcal{R}_m(z)$ to be  proportional to the cosmic star formation rates (\cite{2012ApJ...744...95R}) with a constant time delay  (3.65 Gyr)  between the star formation and
binary black hole merger \cite{2015MNRAS.448.3026W}   and  normalized to  $\mathcal{R}_m(0) = 28\,{\rm Gpc}^{-3}\,{\rm yr}^{-1}$ (see detail in \cite{fan}).  (ii) we  adopt a uniform distribution for  $10 M_{\odot} <M_{1,2} < 30 M_{\odot}$ for $\theta=(M_1,M_2)$, (iii) we  adopt  $(dE/df)_{\rm IMR}$ \cite{2011PhRvD..84h4037A} for the IMR parts of the waveform superimpose $(dE^{\rm SD}/df)_{\rm QNM}$ directly as an additional contribution.

\begin{table}
\begin{tabular}{c|c|c|c}
Network & IMR & SD  & IMR+SD\\ 
\hline\hline
AL+& 7.7436 &  1.0579 &7.9587   \\
 AL+ (100-200)&     0.3637 & 0.6103  &0.9740 \\
Voyager& 54.7418 & 4.4315& 55.2951   \\ 
Voyager (100-200)  & 1.4722  &  2.4326 &    3.9047    \\ 
\end{tabular}
\caption{The network SNR  for {\it fiducial}  IMR alone, spin-down alone, and both combined, for networks combining AL+, Voyager. The first and third lines  show the  optimal SNR.  The  second and forth lines  show  the  SNR  with 100-200 Hz band-pass filter. 
  \label{table}}
\end{table}

\begin{figure}
\includegraphics[width=0.45\textwidth]{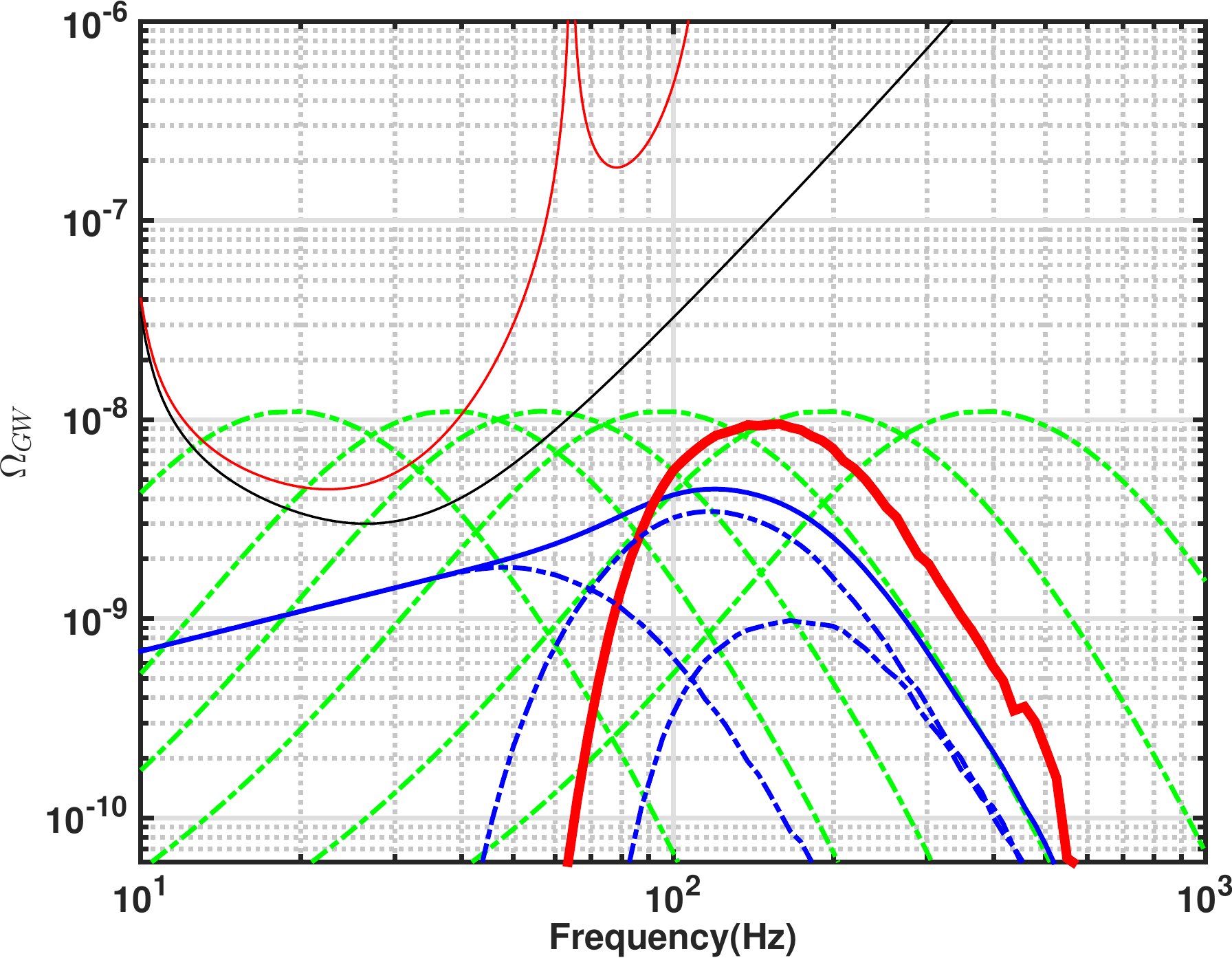}
\caption{We present a set of potential spectra for a BBH
background using the flat  mass distribution model  with the local rate inferred from the O1 and O2 detections. The thick red line represents  the  {\it fiducial} QNM model of the spindown mechanism.  The blue dashed  line represents  the inspiral, merger and ringdown  mechanism and blue line present the total IMR  background.  The alternative models of the spin-down mechanism are shown in green dashed lines assuming different  predominant central  frequency (from let to right are the Parametrized Gaussian Model with $Q=3$ and   $\beta$$= 0.01, 0.0 2, 0.0 3 , 0.05 , 0.1,  0.2 $).  The  thin red  and  black  curve shows the  one year  sensitivity $\Omega_{*}$ of  designed Advanced LIGO network and two co-located and co-aligned  Advanced LIGO like  detectors, respectively (see Eq.~ (12)).  }\label{fig:background}
\end{figure}

 The detection ability of a background  of a detector network also depends on the  overlap reduction function.
 In Fig.~\ref{fig:background}, we plot contributions to $\Omega_{\rm GW}$ from inspiral, merger, ringdown, as well as from spin-down, in comparison with 
\begin{equation}
\Omega_{*} \equiv \frac{S_h^{\rm AdvLIGO} f^3}{ \gamma_{\rm HL}} \sqrt{\frac{1}{2 \Delta f T}}\frac{10\pi^2}{3H_0^2}\,,
\end{equation}
 which sets 1-$\sigma$ sensitivity to $\Omega_{GW}$ in each frequency bin  \cite{2017LRR....20....2R}. Here $\gamma_{\rm HL}$ is the overlap reduction function between the Hanford and Livingston sites of LIGO.  

In Table~\ref{table}, we show the 1-year {\it optimal} SNR for Advanced LIGO+ (first row) and LIGO Voyager (third row), assuming a stochastic background from IMR, SD and IMR+SD, and an optimal filter that corresponds to each case;  the {\it fiducial} QNM model is used.   Even though the SD component does contain more energy than the IMR component, and the emissions are within the detection band of ground-based detectors,  the existence of an addition SD only leads to a small increase in SNR of around $3\%$ --- because $\gamma_{\rm HL}$ significantly decreases above $\sim$50\,Hz. 

\subsection{Separating SD and IMR backgrounds: Fisher Information} 
\label{subsec:extract}

To see that the additional SD background is in fact detectable, we apply a bandpass filter between 100\,Hz and 200\,Hz, and the corresponding SNRs are listed on the second and fourth rows of Table~\ref{table}.   In this band, the gap is more significant.   For LIGO Voyager, the IMR+SD background has a SNR greater than 3, which makes it detectable with greater than 99.7\% confidence, while the SNR for IMR alone is under the 90\% detectability threshold.  

More quantitatively, we can use a Fisher Matrix formalism to obtain the parameter estimation error for the amplitude of the SD background.  Suppose the output of each detector is given by 
\begin{equation}
x_i(f) = n_i(f) + h_i(f)
\end{equation}
where $n_i$ is the noise and $h_i$ the gravitational-wave signal, we construct the correlation between each pair of detectors:
\begin{equation}
z_{ij}(f) = x_i^*(f) x_j(f)  \,.
\end{equation}
 The expectation value of $z_{ij}$ is given by 
\begin{equation}
\langle z_{ij} (f)\rangle = \langle h_i^*  (f)h_j  (f) \rangle =\frac{3H_0^2 T\gamma_{ij} (f) }{20\pi^2 f^3} \Omega^{\rm GW}(f) \equiv c_{ij}(f)
\end{equation}
The covariance matrix is given by 
\begin{eqnarray}
&&\langle z_{ij}^* (f') z_{lm} (f)\rangle  - \langle z_{ij}^* (f') \rangle \langle z_{lm} (f)\rangle  \nonumber\\
&\approx & \langle n_i(f')n^*_j(f') n_l^*(f) n_m(f)\rangle \nonumber\\
& =&\frac{1}{4}( \delta_{il} \delta_{jm}+ \delta_{im} \delta_{jl}) T S_{l} (f) S_m (f) \delta (f-f')
\end{eqnarray}
In fact, we only need to consider $z_{ij}$ with $i>j$, which means the $z_{ij}(f)$'s all have independent noise, and the likelihood function for a particular $z_{ij}(f)$ is given by~\cite{Romano2017} 
\begin{equation}
p[z_{ij}(f) | S_h(f)] \propto \exp \left[-\int_{0}^{+\infty}  \frac{4|z_{ij}(f) - c_{ij}(f)|^2}{TS_{i}(f) S_j(f) }df\right]
\end{equation}
If we only consider the SNR, we will simply sum all the frequencies and detector pairs by quadrature, and take the square root, and obtain
\begin{equation}
\mbox{SNR}= \left[\sum_{i>j} 8\int_0^{+\infty} df \frac{c_{ij}^2}{T S_i (f)S_j (f)} \right]^{1/2}
\end{equation}
which agrees with Eq.~\eqref{eq:snrCC}.   Suppose $\Omega_{\rm GW}$ depends on a set of parameters $\theta^\alpha$, then we can obtain the Fisher matrix
\begin{equation}
\Gamma_{\alpha\beta} = \left(\frac{3H_0^2 \sqrt{2T}}{10\pi^2}\right)^2\sum_{i>j}\int_0^{+\infty} df \frac{\gamma_{ij}^2
}{f^6 S_i^h S_j^h}
\frac{\partial \Omega^{\rm GW}}{\partial \theta^\alpha}
\frac{\partial \Omega^{\rm GW}}{\partial \theta^\beta}\,.
\end{equation}
Suppose we have a simple model with 
\begin{equation}
\Omega^{\rm GW}= \sum_J \alpha_J \Omega_J
\end{equation}
where in our case $J$ is for IMR and SD.  We obtain
\begin{equation}
\Gamma_{JK} = \left(\frac{3H_0^2 \sqrt{2T}}{10\pi^2}\right)^2\sum_{i>j}\int_0^{+\infty} df \frac{\gamma_{ij}^2 \Omega_J \Omega_K
}{f^6 S_i^h S_j^h}
\end{equation}
and the {standard deviation of the estimation on  $\alpha_{\rm SD}$ } is given by 
\begin{equation}
\label{alphaSD}
\sigma_{\alpha_{\rm SD}}= \left(\Gamma_{\rm SD,SD} -\frac{\Gamma_{\rm SD,IMR}^2}{\Gamma_{\rm IMR,IMR}}\right)^{-1/2}
\end{equation}

In the special (optimistic) case that the SD and IMR spectra take very different shapes, the correlation term in Eq.~\eqref{alphaSD} can be ignored, and we recover
\begin{equation}
\label{alphaSDapp}
\sigma_{\alpha_{\rm SD}} \approx  \Gamma_{\rm SD,SD}^{-1/2} = \sqrt{1/{\rm SNR}^2}
\end{equation}

\begin{figure}
\includegraphics[width=0.45\textwidth]{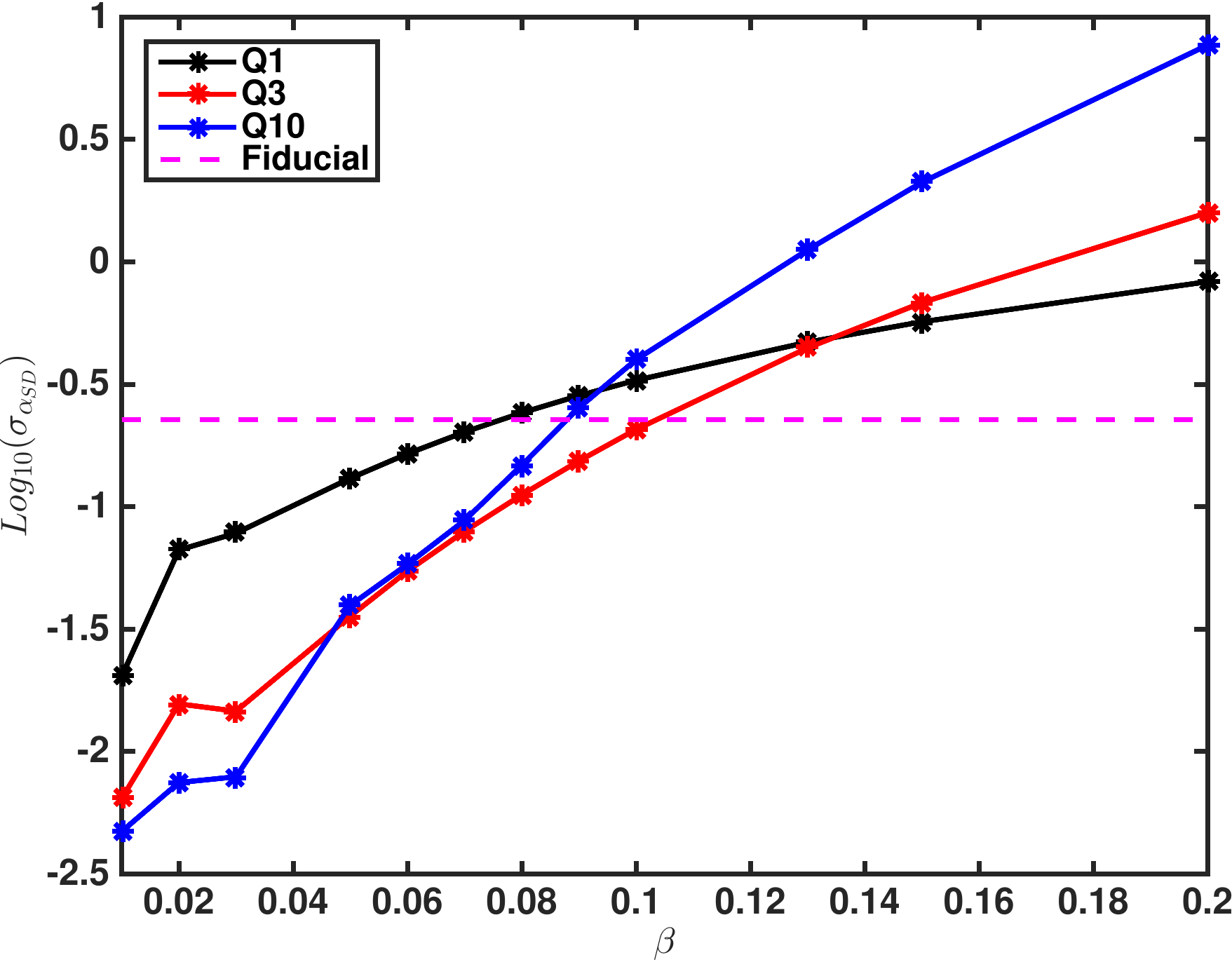}
\caption{ The  standard deviation  $\sigma_{\alpha_{\rm SD}}$  for one year Voyager  observation for SD models. Q1, Q3 and Q10 represent  Parametrized Gaussian Model with quality factor  Q=1, 3 and 10, respectively. \label{fig:error}}
\end{figure}

\subsection{Results}
\label{subsec:results}
In Fig.~\ref{fig:error}, we plot $\sigma_{\alpha_{\rm SD}}$ for Parametrized Gaussian Models with different values of $\beta$ and $Q$ (black, red and blue curves), and  the fiducial model (magenta dashed line).   

First of all, the fiducial model has   $ \sigma_{\alpha_{\rm SD}}= 0.22$. This is rather well approximated by Eq.~\eqref{alphaSDapp} (and SNR from Table. \ref{table}), because the SD background in this case has a rather different shape from the IMR spectrum, as shown in Fig.~\ref{fig:background}.

As for the Parametrized Gaussian Models, for cases with $Q=1,3,$ and 10, Fig.~\ref{fig:error} indicates that the SD background can be extracted if $\alpha_{\rm SD}$ is greater than 0.3, when $\beta$ is no larger than 0.1.   In other words, for $\beta \le 0.1 $, the background is detectable if spin-down carries away more than 30\% of spin energy.  The better sensitivity to $\alpha_{\rm SD}$ for smaller $\beta$  can be attributed to the fact that overlap-reduction function seriously limits the detectability of high-frequency backgrounds.   

%
%

\section{Conclusions and Discussions}  
\label{sec:conclusion}

In this paper, we  argued that spinning black holes, or ultracompact objects, can spin down  without being inconsistent with LIGO observations --- as long as the spin-down time scale is much longer than the dynamical time scales of the black holes.   We have also noted that spinning black holes (or ultracompact objects) do form from binary black-hole coalescence, as indicated by LIGO observations,  and that they do carry  significant amount of spin energy ---  more than what is emitted during the entire coalescence,  as seen in Fig.~\ref{fig:energy}.

 This  BH spin-down stochastic background can be an interesting target of  next-generation GW detector networks, such as the LIGO Voyager.  As has been estimated by our Fisher Matrix approach,  both for the fiducial model and the $\beta <0.1$ case of the Parametrized Gaussian Model, the standard deviation for estimating the magnitude of the SD backgrounds is $<30\%$ for  a 1-year observation of Voyager.  Beyond the detection (see Ref.~\cite{2017LRR....20....2R} for review),  to  extract the  parameters associated with different stochastic gravitational-wave background models,  a Bayesian approach is  being developed within the GW community, such as described in Refs.~\cite{2012PhRvL.109q1102M, 2017arXiv171200688S}.  In future studies we hope to investigate such a Bayesian application to the detection of the SD backgrounds.  

Even though a SD stochastic gravitational-wave background is not detected by future detectors, we can still use the null result to put constraints on the existence and nature of emission, therefore shedding light on black-hole superradiance.  For example, the $\sigma_{\alpha_{\rm SD}}$ shown in Fig.~\ref{fig:error} characterizes the magnitude of upper limit we can pose on $\alpha_{\rm SD}$ for each $\beta$ and $Q$.

 \begin{acknowledgments}

We thank  A. Matas and Xingjiang Zhu for valuable comments. X. F.  is  supported by  Natural Science Foundation of China under Grants (No. 11633001,  No. 11673008) and  Newton International Fellowship Alumni Follow on Funding.  Y. C. is supported by US NSF Grants PHY-1708212,  PHY-1404569 and PHY-1708213.
 \end{acknowledgments}


\bibliography{PRD_re_sub_re_0728.bbll}
\end{document}